\documentclass[runningheads]{llncs}

\usepackage[T1]{fontenc}
\usepackage{graphicx} 
\usepackage[ruled,linesnumbered]{algorithm2e}
\usepackage{graphicx}
\usepackage{soul}
\usepackage{xcolor}
\begin{document}
\title{Aid Nexus : A Blockchain Based Financial Distribution System}

\author{Md. Raisul Hasan Shahrukh\inst{1} \and Md. Tabassinur Rahman\inst{2} \and Nafees Mansoor \inst{3}}

\institute{Department of Computer Science \& Engineering, \\ University of Liberal Arts Bangladesh, Dhaka-1207, Bangladesh \\
\email{raisul.hasan.cse@ulab.edu.bd}\inst{1}, \email{tabassenur.rahman.cse@ulab.edu.bd}\inst{2}, \email{nafees@ieee.org}\inst{3}
}

%

%
%
%
\maketitle              
\begin{abstract}
Blockchain technology has emerged as a disruptive force with transformative potential across numerous industries, promising efficient and automated solutions that can revolutionize traditional systems. By leveraging decentralized ledger systems, blockchain offers enhanced security, transparency, and transaction verification without the need for intermediaries. The finance sector is exploring blockchain-based solutions for payments, remittances, lending, and investments, while healthcare adopts the technology for medical record keeping, supply chain tracking, and data management. Similarly, supply chain management benefits from blockchain's ability to enhance transparency, traceability, and accountability from raw materials to finished products. Other sectors, including real estate, energy, and government, are also investigating blockchain-based solutions to improve efficiency, security, and transparency. Furthermore, smart contracts within the blockchain enable process automation, reducing manual intervention in distribution workflows. AidNeux, a consortium-based blockchain DApp, reimagines the distribution of financial assistance by addressing inefficiencies and opaqueness. Using smart contracts ensures the security and directness of money transfers. Its robust digital identity verification and real-time auditability reduce fraud risks and strengthen accountability, thereby presenting a scalable, transparent solution to problems inherent to conventional financial aid systems. 
\end{abstract}

\section{Introduction}
At the advent of the 21st century, unprecedented technological advancements reshaped the traditional landscape in numerous industries. Blockchain is a disruptive, revolutionary technology that is widely recognized for its disruptive potential. In 2008, the pseudonymous entity Satoshi Nakamoto introduced the digital currency Bitcoin using Blockchain technology. Since then, the evolution of Blockchain technology has expanded beyond cryptocurrencies and is poised to revolutionize a vast array of industries, with a particular impact anticipated in the financial sector. This eliminates the need for centralized authorities or intermediaries \cite{nb1}. The decentralized, immutable, and transparent nature of blockchain technology promises to streamline processes, improve efficiency and security, and cultivate confidence in the financial sector. Moreover, the transformative potential of blockchain also reveals the development of an innovative on-demand, lightweight commodity delivery system designed to bring unprecedented levels of efficiency and transparency to the logistics industry \cite{nb2}.

The fundamental components of blockchain technology are the data structure, peer-to-peer network, and consensus mechanisms \cite{nb3}. Data structures organize transactions into blocks, each with a unique identifier and a hash of the previous block, thereby ensuring the integrity and security of the data \cite{nb4}. Peer-to-peer networks enable direct communication between nodes, thereby augmenting robustness and efficiency \cite{nb5,nb6}. Proof-of-Work (PoW), Proof-of-Stake (PoS), and Practical Byzantine Fault Tolerance (PBFT) are consensus mechanisms that validate transactions and determine the blockchain's state \cite{nb7}. There are three types of blockchain technologies: public, private, and consortium. Public blockchains, such as Bitcoin and Ethereum, allow for universal participation but confront scalability, speed, and privacy issues \cite{nb8}. Private blockchains, which are under the control of a specific group, offer enhanced efficiency and privacy at the expense of decentralization \cite{nb9}. Consortium blockchains provide a balance by enabling a predefined group of nodes to maintain decentralization while improving privacy and control \cite{nb10}.Due to its security, transparency, and efficiency, blockchain technology possesses immense transformative potential in the financial sector. As demonstrated in the paper \cite{nb11}, it can improve cross-border transactions by reducing complexity and costs while increasing efficiency and security.

The paper \cite{nb12} explored how blockchain can expedite trading and stock exchange platforms by eliminating intermediaries. Smart contracts, facilitated by blockchain technology, automate the execution of agreements, reducing enforcement costs and boosting productivity \cite{nb13}. The immutability and transparency of blockchain can enhance the auditability of financial transactions by providing a secure audit trail and streamlining regulatory compliance \cite{nb14}. The decentralized nature of blockchain offers the financial sector optimistic data security and fraud detection prospects \cite{nb15}. As demonstrated by Purohit et al. \cite{nb16}, consortium blockchains that offer a balance between complete transparency and restricted access have the potential to satisfy the privacy requirements of the financial industry.

AidNeux is an innovative decentralized application (DApp) utilizing consortium blockchain technology to revolutionize traditional pension systems by addressing prevalent issues such as inefficiency, lack of transparency, and administrative complexities. AidNeux's implementation of automated smart contracts that dictate the pension amount and delivery schedule for eligible beneficiaries, thereby eliminating the need for intermediaries and enhancing transparency and efficiency, is a defining characteristic. In addition, AidNeux includes a trustworthy digital identity verification system that validates pension recipients' eligibility, utilizing the immutability of blockchain for secure storage and authentication of identities. This DApp guarantees the direct transfer of pension funds from authorities to beneficiaries, thereby reducing administrative costs and corruption risks. Its transparency enables real-time auditability of fund distribution, thereby enhancing accountability and decreasing the likelihood of fraud. AidNeux exemplifies a revamped pension fund distribution strategy with its superior scalability and interoperability, effectively leveraging consortium blockchain's unique advantages to surmount traditional system challenges.

The purpose of this paper is to provide an overview of the potential transformative effects of blockchain technology in the financial sector. Section 2 examines the issue of financial distribution in humanitarian sectors. The purpose of Section 3 is to elucidate the fundamental principles of the proposed system, while Section 4 demonstrates the design and implementation of this system. The paper concludes with a summary of the blockchain revolution and its future in Section 5. 

\section{Related Works}
Blockchain technology is highly transformative, especially in the financial sector, as it offers increased efficiency, decreased costs, and enhanced security \cite{nb7}. This technology provides a more efficient, secure, and cost-effective method for financial transactions by eliminating the need for intermediaries and enhancing auditability by virtue of its inherent immutability and transparency \cite{nb8}. The financial industry increasingly recognizes the value of consortium blockchain, a partially private variant managed by a predetermined group of nodes, for its balance of transparency and privacy. \cite{nb12,nb17} This variant has numerous applications. The authors \cite{nb11} propose an intelligent cross-border transaction system based on consortium blockchain, demonstrating the technology's applicability outside of the financial sector. 

The paper \cite{nb12} suggests consortium blockchain-based decentralized stock exchange platforms can mitigate traditional stock market problems like high operational costs, lack of transparency, and susceptibility to fraud. However, consortium blockchains' limitations, such as scalability and interoperability with conventional stock exchanges, may present obstacles. Future research should concentrate on these issues and investigate possible innovative implementations in industry \cite{nb12}. The researchers \cite{nb16} advocate for the blockchain consortium system HonestChain for secure data sharing in health information systems. The system incorporates the benefits of both public and private blockchains to circumvent their respective limitations, thereby providing enhanced security, privacy, and regulatory compliance for the data sharing of health information systems. The method, which includes attribute-based encryption, access control mechanisms, and decentralized authorization, enables precise regulation of access to confidential health information \cite{nb16}. 

The Authors \cite{nb16} indicate the potential of consortium blockchain for improved data privacy and security in health information systems, highlighting its utility for financial institutions requiring secure data sharing. Despite its infancy, research into blockchain implementations in various industries, including pension services \cite{nb18}, is promising. Regulatory constraints, interoperability concerns, and user adoption are obstacles, but the potential benefits merit further investigation \cite{nb11,nb19}. This study investigates Vehicular Ad Hoc Networks (VANETs) - networks of interconnected vehicles that self-organize. These networks enable substantial improvements in vehicular communication, transportation security, and overall traffic management effectiveness \cite{nb20,nb21}. Moreover, consortium blockchain enables substantial advances in data sharing and storage in vehicular ad hoc networks (VANETs) \cite{nb15}. Currently, security risks, trust issues, and storage constraints impede VANET data sharing. Consortium blockchain addresses these obstacles by providing a secure and transparent platform \cite{nb15}. 

However, the ephemeral character of VANETs and their limited computational resources pose challenges for consortium blockchain implementation. Researchers propose the use of lightweight consensus algorithms, privacy-preserving mechanisms, and efficient data storage techniques to mitigate these issues \cite{nb15}. To thoroughly comprehend the potential and limitations of consortium blockchain in this context, additional research is required. Despite its limitations, consortium blockchain technology is emerging as a significant instrument in industries including VANETs, cross-border transactions, and logistics. For example, a research paper \cite{nb17} highlights the advantages of consortium blockchain for secure data interchange and the provision of customized services in intelligent transportation systems. The introduction of blockchain technology has catalyzed remarkable transformations in numerous industries, resulting in increased efficiency, security, and cost-effectiveness. The financial sector has shown particular interest in consortium blockchain technology due to its combination of transparency and anonymity \cite{nb12}. 

Consortium blockchain combines features of both public blockchain and private   blockchain, providing a secure, efficient, and dependable platform for data exchange and service delivery. In intelligent transportation systems, for instance, the research  \cite{nb12} argue that a consortium blockchain-based system outperforms conventional centralized systems, which are more prone to data manipulation and privacy violations. Smart household systems also have promising implications for consortium blockchain. The authors, in the research paper \cite{nb22} proposes a consortium blockchain system for data privacy protection utilizing homomorphic encryption, which enables the processing of encrypted data without decryption. Contextually, consortium blockchain is proposed as a decentralized, transparent, and secure solution for mobile malware detection \cite{nb23}. The system enables collaborative detection and prevention of malware infections across a network of mobile devices. The consortium blockchain has the potential to improve smart city governance \cite{nb8}. Proposed is a public participation consortium blockchain framework to enhance smart city decision-making and citizen participation.

The paper \cite{nb24} demonstrates that the healthcare industry, which is plagued by data security and access concerns, can benefit substantially from consortium blockchain. The proposed system satisfies the disparate data access requirements of various stakeholders and enhances the efficacy and security of medical data management.
Additionally, consortium blockchain can improve safety and effectiveness in high-risk industries, such as coal extraction \cite{nb25}. The proposed solution encourages data sharing, collaboration, and accountability among stakeholders, resulting in a reduction in safety incidents. In addition, consortium blockchain systems can improve data accountability and provenance monitoring \cite{nb14}, which has significant implications for data management in the healthcare, finance, and government sectors. In data management for smart grids, consortium blockchain can surmount obstacles such as a lack of transparency and trust \cite{nb14}. In the pension system, consortium blockchain technology proposes solutions to increase effectiveness and transparency, enabling secure pension access and administration \cite{nb26}. In addition, the combination of consortium blockchain technology and attribute-based signatures can enhance the privacy protection of Vehicle-to-Grid (V2G) networks \cite{nb18}. Similar methods, such as multi-party computation and homomorphic encryption, can be utilized to secure smart contracts \cite{nb13}. A consortium blockchain can also safeguard peer-to-peer file storage and sharing \cite{nb27}, whereas a trust-based hierarchical consensus mechanism cannot.

In conclusion, consortium blockchain technology has enormous potential in a variety of industries. It offers solutions to problems with which conventional centralized systems have labored and indicates promising future developments in these fields.

\section{Proposed System: Aid Nexus}
This section describes the Aid Nexus platform, which aims to improve the transparency and efficacy of humanitarian aid distribution. The design, which is represented by a block diagram and Unified Modeling Language (UML) models, illustrates the system's essential components, their interactions, and object-oriented characteristics. These provide a clear comprehension of the structure and functionality of the system, laying the groundwork for future development and implementation.

\subsection{System Overview}
The AidNeux system, shown in Fig. 1, operates within a consortium blockchain framework that includes multiple entities, including a fund-distributing entity, eligible pension recipients, and potential financial service providers. The principal node, which is administered by the organizing entity, oversees the system, including the upkeep of smart contracts, beneficiary registration, fund additions, and distribution. Pension-eligible recipients represent recipient nodes, which are authorized by the organization node to interact with the smart contract in order to verify their account balance or acquire funds. On the Consortium Blockchain Network, transactions initiated by the organization node, such as fund additions or distributions, are recorded. Banks and other financial service providers could interact with the smart contract to enable off-chain transactions. When the smart contract is deployed, the recipient's banking information is embedded. During pension distribution, an authorized organization node verifies the status and balance of the recipient and records the transaction.
\begin{figure}[htp]
    \centering
    \includegraphics[width=\linewidth]{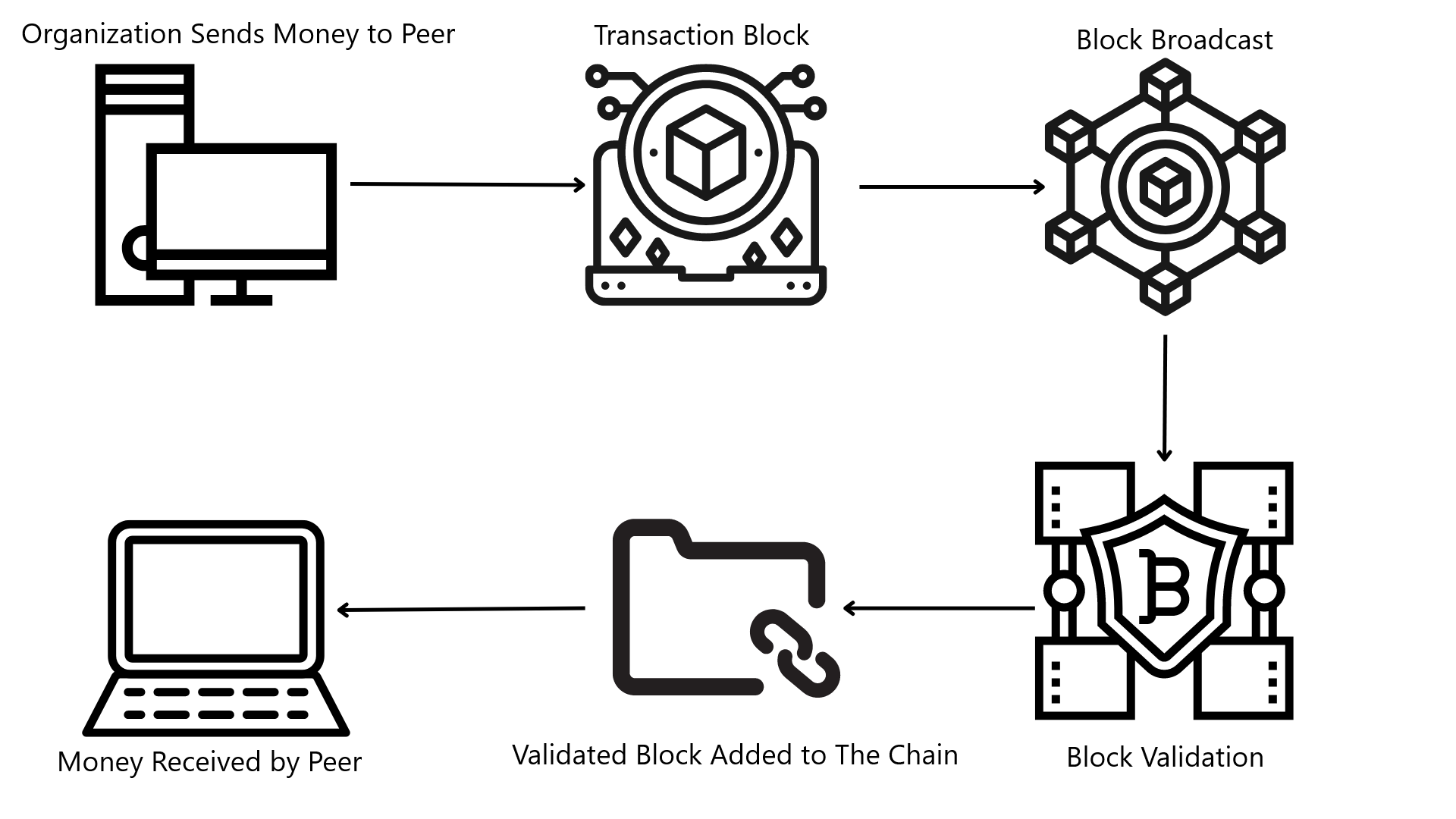}
    \caption{Aid Nexus Block Diagram}
    \label{fig:Aid Nexus Block Diagram}
\end{figure}
 Recipient nodes can validate their balances, and financial service providers facilitate off-chain transactions by retrieving data from the blockchain. All transactions are recorded openly, ensuring security, auditability, and traceability, and the consortium-based structure improves efficiency and control.

\subsection{System Illustrations}
The use case diagram for AidNeux's system, shown in Fig. 2, vividly depicts the interaction between the various entities and the system itself, providing a concise illustration of the overall functionality and flow.
\begin{figure}[htp]
    \centering
    \includegraphics[width=\linewidth]{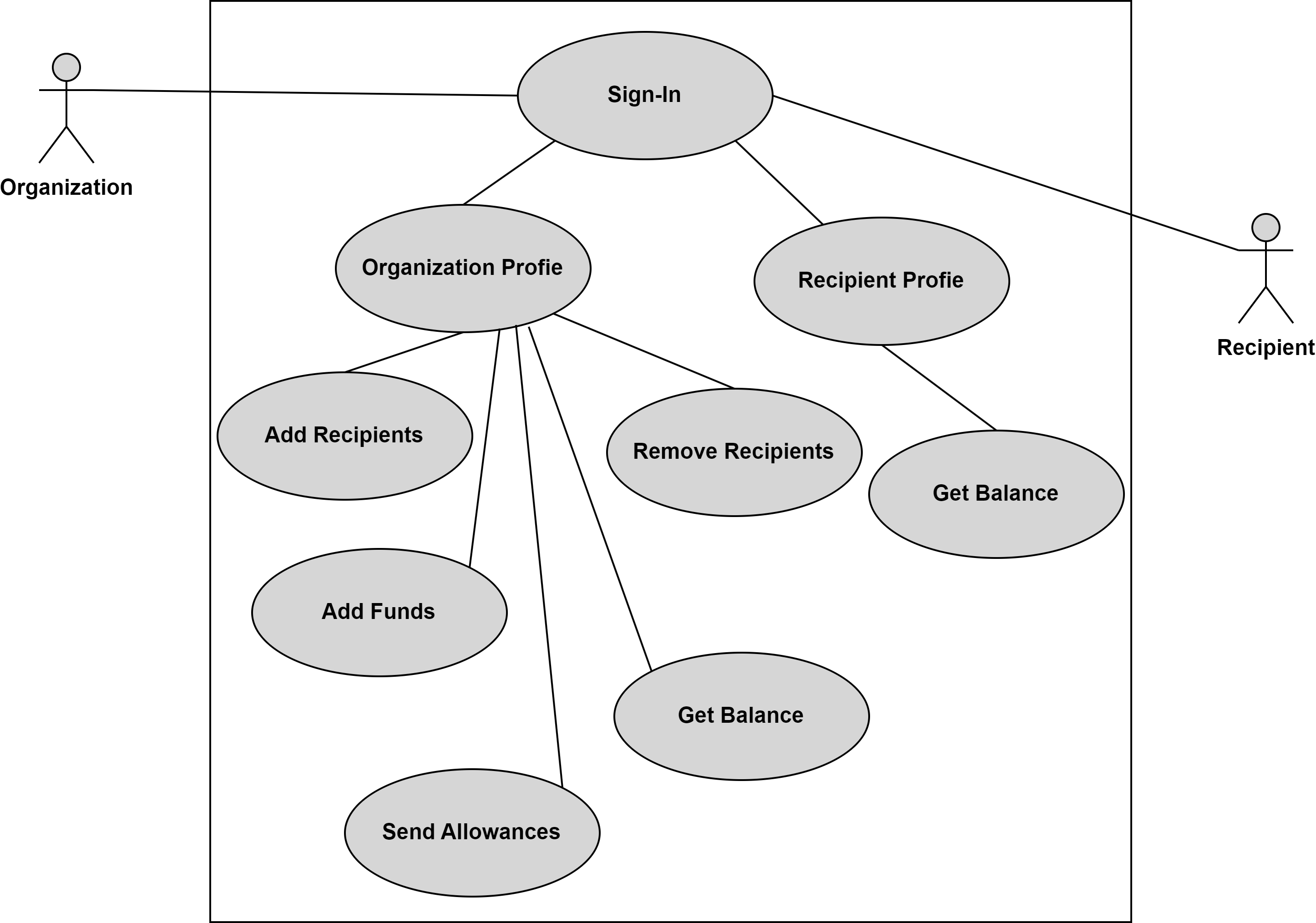}
    \caption{Aid Nexus Use Case Diagram }
    \label{fig:Aid Nexus Use Case Diagram }
\end{figure}
The 'Organization' is depicted as the primary protagonist who carries out the essential duties. These responsibilities consist of adding and removing recipients, adding funds, sending allowances, and registering bank accounts. 
It is evident from this interaction that the organization possesses the controlling power and serves as the system's backbone.
The 'Recipient' is an additional important participant in this system. Although the recipient does not directly interact with the system, their interaction is implied because they are the recipients of the organization's payments. In addition, the diagram clearly delineates responsibilities and interactions, indicating that the system is well-designed to prevent confusion.
With its emphasis on illustrating how various entities interact with the system, the use case diagram [Fig. 2] concisely captures the essence of AidNeux's operation. It provides an excellent method for visually communicating complex processes and interactions in a way that is easily understood. The interaction between these actors and the system exposes a robust and well-structured approach to aid distribution, which lays the groundwork for an effective, transparent, and secure distribution of resources.

The activity diagram of the AidNeux system, shown in Fig. 3, illustrates the sequential sequence of actions that occur within the system. The 'Organization', which has the authority to add funds to the system, is the starting point. Once the funds have been effectively added, the organization will manage recipients by adding or removing them based on specific criteria. Evidently, recipient administration is essential for maintaining the integrity and equity of the aid distribution process. After properly managing the beneficiaries, the organization can then register the bank accounts of the authorized recipients. This phase is essential for ensuring that funds are transferred to the correct accounts, highlighting the system's emphasis on security and precision.
Finally, the organization is able to provide recipients with allowances. This action is dependent on the successful completion of the preceding stages, indicating a well-coordinated and efficient procedure. Therefore, the activity diagram [Fig. 3] highlights the sequence of operations within the AidNeux system, illustrating how each action is a prerequisite for the next and contributing to the system's overall effectiveness and dependability.
\begin{figure}[htp]
    \centering
    \includegraphics[width=\linewidth]{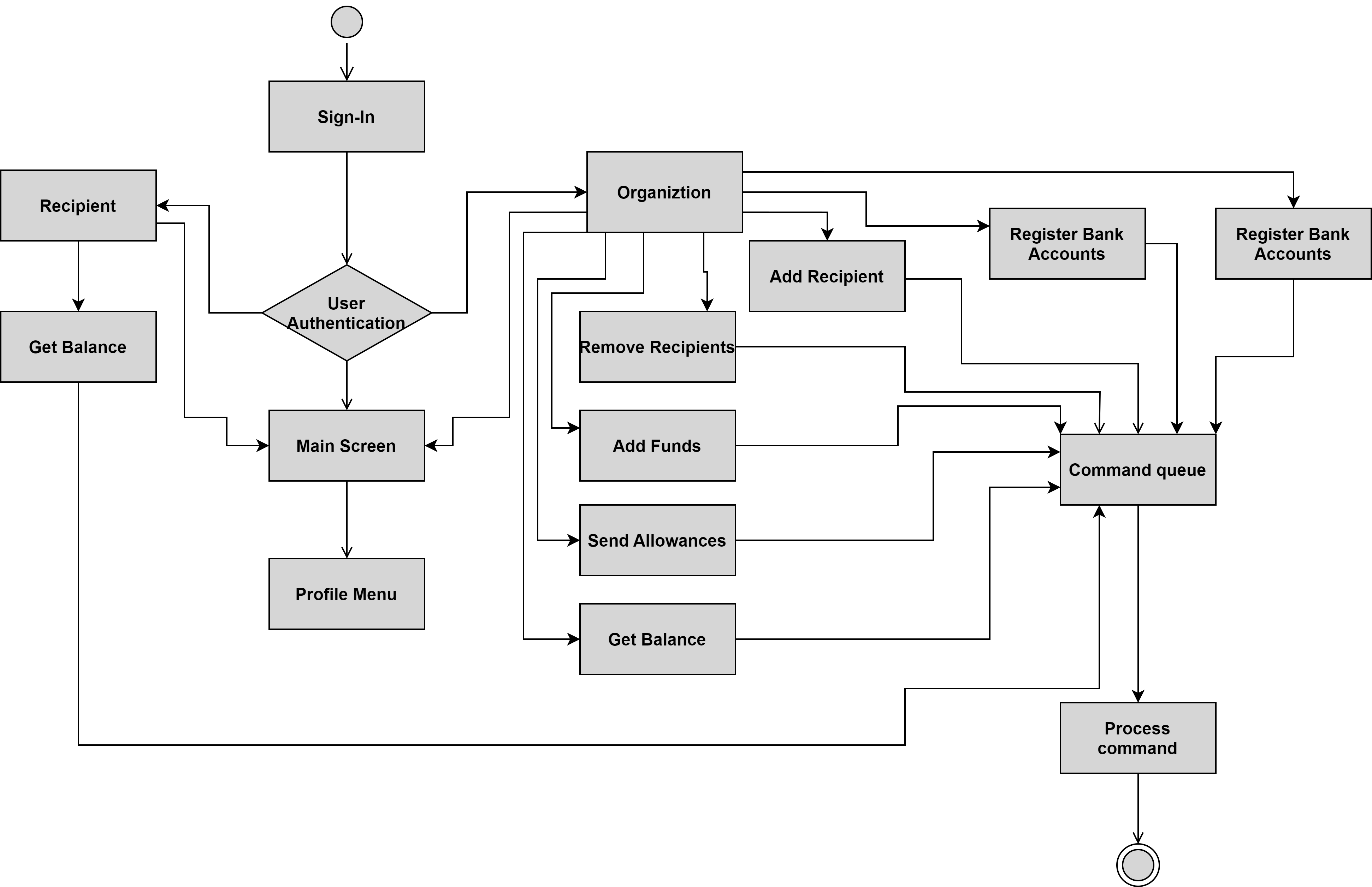}
    \caption{Aid Nexus Activity Diagram }
    \label{fig:Aid Nexus Activity Diagram }
\end{figure}
\newline

The class diagram of the AidNeux system, shown in Fig. 4, depicts the structural relationships and dependencies between the system's various entities. The central component of the system is the 'aNp' class, which represents the smart contract. This class incorporates several crucial properties, including 'organization', 'recipients', 'balances', and 'bankAccounts'. Each attribute serves a distinct purpose, delineating the smart contract's properties and its interaction with the system.
\begin{figure}[htp]
    \centering
    \includegraphics[width=\linewidth]{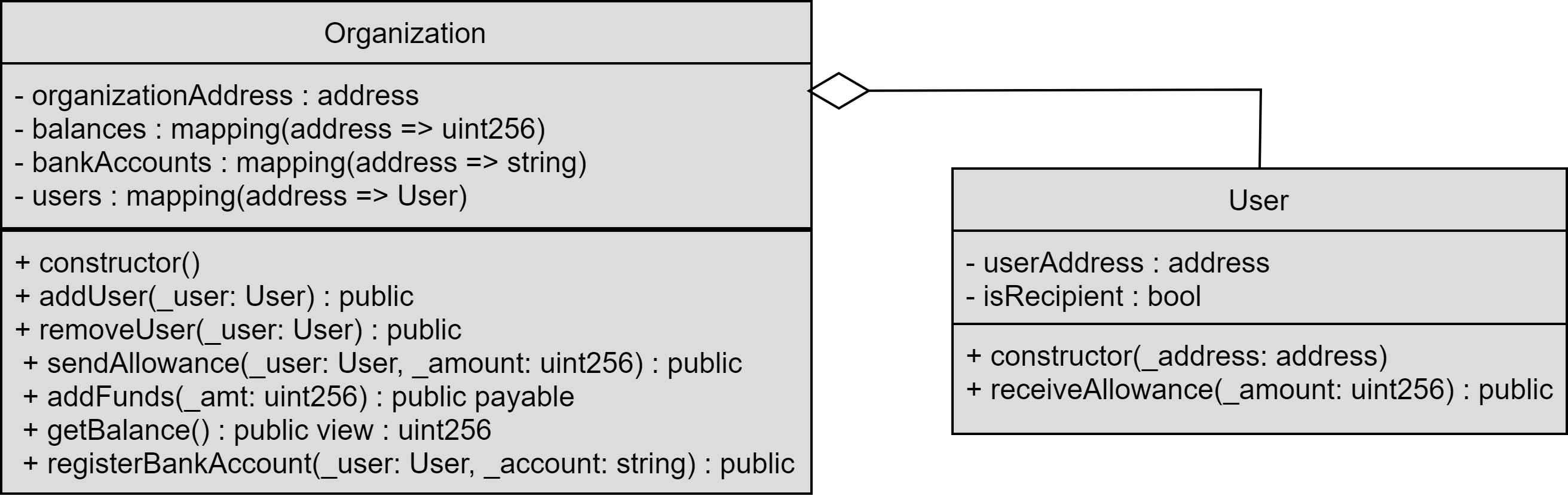}
    \caption{Aid Nexus Class Diagram  }
    \label{fig:Aid Nexus Class Diagram }
\end{figure}
The 'organization' attribute is in charge of administering the entire system, as it has exclusive permission to invoke vital methods within the class. The 'Recipients' attribute denotes the addresses of individuals or organizations that are eligible to receive the funds. The 'organization' tightly controls their administration, including additions and removals.
The 'balances' attribute monitors and maintains the amounts designated to each address. It ensures transparent monitoring of funds, thereby enhancing the system's integrity. In contrast, 'BankAccounts' stores the recipient's bank account information, which is essential for facilitating transactions.
The methods defined within the class, such as 'addRecipient', 'removeRecipient',
'sendAllowance', 'addFunds', 'getBalance', and 'registerBankAccount', serve distinct purposes to administer the overall system operations.
Essentially, the class diagram [Fig. 4] offers valuable insight into the inner workings of the AidNeux system by providing a distinct perspective on the system's structure and the interaction between its various components.

\section{Development \& Implementation of the Proposed System}
The AidNeux architecture, constructed on a consortium blockchain, is centered on a Solidity-based smart contract, 'aNp', which administers transactions and node interactions. The 'organization', the contract holder, supervises the distribution of funds. It is responsible for deploying the contract, managing recipient information with the 'addRecipient', 'removeRecipient', and 'registerBankAccount' functions, and distributing funds with the 'sendAllowance' function. Beneficiaries of the fund ('recipients') can interact with the contract via the 'getBalance' function, but control over the funds is restricted until disbursement. 

Compared to public blockchains, the consortium blockchain network provides a decentralized, transparent platform for transaction records, thereby enhancing security and auditability. Potential external parties, such as institutions, could interact with the system to conduct off-chain fund transfers. These entities would extract transaction information from the blockchain contract and implement transactions through the conventional banking system. This architecture combines the functionality of smart contracts to provide a secure, efficient, and transparent platform for pension fund distribution.

\subsection{Smart Contract}
The aNp (aidnexus) contract, crafted in Solidity, is an adept system for regulated financial distribution within a consortium blockchain. It's designed to manage, secure, and efficiently distribute financial allowances under an authoritative organization. The contract facilitates flexible recipient management, allowing additions or removals solely by the organization. The core function permits the organization to distribute allowances, employing strict rules to verify recipient authorization and check sufficient funds, ensuring secure transactions. The contract further allows the organization to add funds, promoting operational continuity. 

It also links recipient addresses with bank accounts using secure keccak256 encryption, protecting sensitive data while enabling efficient financial transfers. In essence, the aNp contract leverages blockchain technology for secure, transparent, and efficient financial distribution.

\begin{algorithm}[H]
\LinesNotNumbered
\SetAlgoLined
\KwData{msg.sender as the current function caller.}
\KwResult{Initialize the contract with msg.sender as the organization.}
    organization = msg.sender\;
\caption{Constructor}
\end{algorithm}

The constructor function [Algorithm 1] initializes the contract when it's deployed on the blockchain. In the context of the aNp contract, the constructor function assigns the contract deployer's address (represented as msg.sender in Solidity) as the "organization". The organization is a key actor in this contract, authorized to execute critical functions such as adding or removing recipients, sending allowances, adding funds, and registering bank accounts.

\begin{algorithm}[H]
\LinesNotNumbered
\SetAlgoLined
\KwData{msg.sender as the current function caller, \_recipient as the address of the new recipient.}
\KwResult{Add a new recipient.}
 \If{msg.sender == organization}{
    recipients[\_recipient] = true\;
 }
\caption{AddRecipient}
\end{algorithm}

The AddRecipient function [Algorithm 2] is essential for expanding the scope of the organization's financial distribution. It allows the organization to authorize new recipients, who can then receive allowances. This function ensures that only the organization can add recipients, enforcing a layer of access control and preventing unauthorized entities from manipulating the list of recipients.

\begin{algorithm}[H]
\LinesNotNumbered
\SetAlgoLined
\KwData{msg.sender as the current function caller, \_recipient as the address of the recipient to be removed.}
\KwResult{Remove a recipient.}
 \If{msg.sender == organization}{
    recipients[\_recipient] = false\;
 }
\caption{RemoveRecipient}
\end{algorithm}

The RemoveRecipient function [Algorithm 3] provides the organization with the ability to revoke the recipient status of an address. This feature is necessary for managing changes in the system, such as when a recipient is no longer eligible or when there's a need to prevent misuse. Like the AddRecipient function, this function can only be called by the organization, ensuring a secure and controlled system.

\begin{algorithm}[H]
\LinesNotNumbered
\SetAlgoLined
\KwData{msg.sender as the current function caller, \_recipient as the address of the recipient, \_amount as the amount to be sent.}
\KwResult{Send allowance to a recipient.}
 \If{msg.sender == organization AND recipients[\_recipient] == true AND balances[organization] >= \_amount}{
    balances[organization] -= \_amount\;
    emit AllowanceSent(\_recipient, \_amount)\;
 }
\caption{SendAllowance}
\end{algorithm}

The SendAllowance function [Algorithm 4] enables the transfer of funds from the organization to an authorized recipient. Before transferring, the function verifies that the recipient is authorized and that the organization has enough funds to cover the transfer. This robust verification procedure ensures the system's financial integrity and enforces the rules set by the organization.

\begin{algorithm}[H]
\LinesNotNumbered
\SetAlgoLined
\KwData{msg.sender as the current function caller, \_amt as the amount to be added.}
\KwResult{Add funds to the organization's balance.}
 \If{msg.sender == organization}{
    balances[organization] += \_amt\;
    emit FundsAdded(msg.value)\;
 }
\caption{AddFunds}
\end{algorithm}

The AddFunds function [Algorithm 5] allows the organization to increase its balance. The added amount can be further utilized to distribute allowances to the recipients. This function underpins the contract's operational continuity by ensuring that the organization can add funds whenever necessary. Moreover, it records the transaction on the blockchain, providing a traceable financial record.

\begin{algorithm}[H]
\LinesNotNumbered
\SetAlgoLined
\KwData{msg.sender as the current function caller.}
\KwResult{Return the balance of the function caller.}
 return balances[msg.sender]\;
\caption{GetBalance}
\end{algorithm}

The GetBalance function [Algorithm 6] is a simple yet essential feature of the contract that allows any address to check its current balance. This feature promotes transparency and allows the organization and recipients to monitor their fund levels.

\begin{algorithm}[H]
\footnotesize
\LinesNotNumbered
\SetAlgoLined
\KwData{msg.sender: Current function caller}
\KwData{\_recipient: Recipient address}
\KwData{\_account: Account to be registered}
\KwResult{A registered bank account for a recipient.}
 \If{msg.sender == organization AND recipients[\_recipient] == true AND bytes(\_account).length > 0}{
    bankAccounts[\_recipient] = keccak256(abi.encodePacked(\_account))\;
    emit BankAccountRegistered(\_recipient, bankAccounts[\_recipient])\;
 }
\caption{RegisterBankAccount}
\end{algorithm}

The RegisterBankAccount function [Algorithm 7] facilitates the association of bank account details with recipient addresses. It incorporates Keccak256 encryption for storing bank account details, ensuring the privacy and security of this sensitive information. The organization is responsible for registering bank accounts, further enhancing system security by centralizing this process.

\subsection{Aid Nexus User Interface(UI)}
This section introduces the Aid Nexus implementation phase, with an emphasis on the user interface (UI) of the application. It depicts how the designed components interact with one another in a functional context. The AidNeux system interface, shown in Fig. 5, will likely be an intuitive, user-friendly graphical user interface (GUI), facilitating access and operation for both organizations and recipients. Through input fields corresponding to smart contract functions such as addRecipient, removeRecipient, registerBankAccount, and sendAllowance, the organization can manage recipients, register bank accounts, and distribute funds. 
\begin{figure}[htp]
    \centering
    \includegraphics[width=\linewidth]{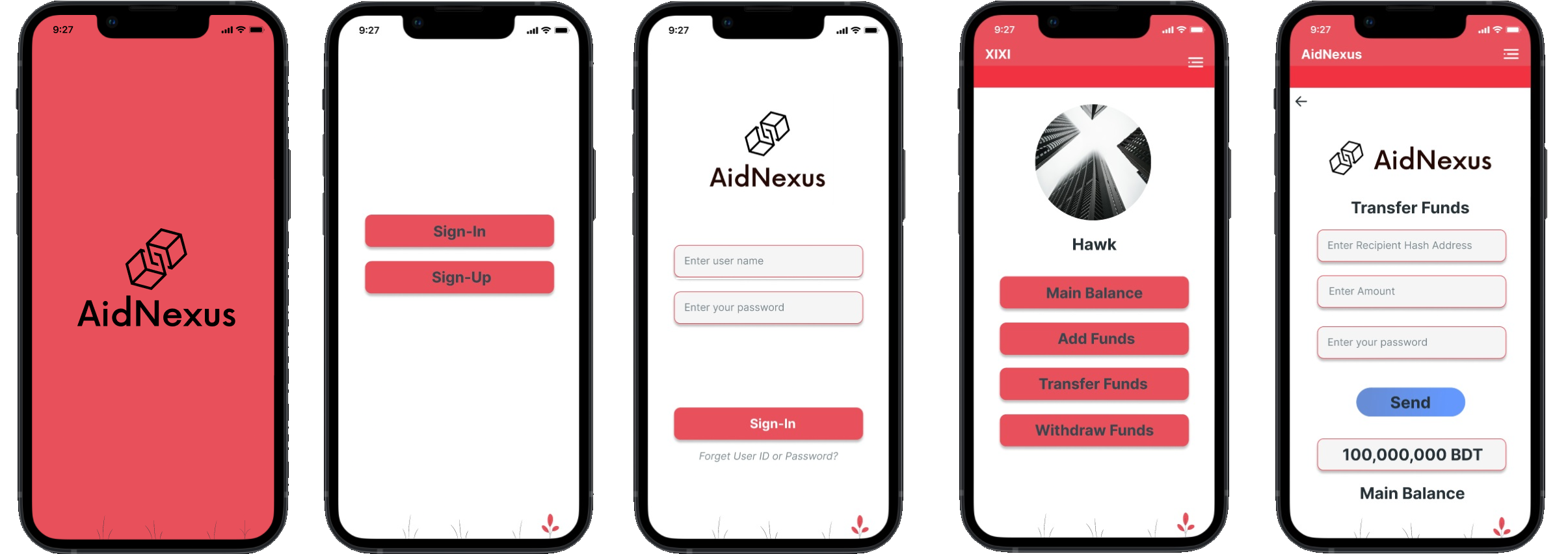}
    \caption{Aid Nexus System Interface }
    \label{fig:Aid Nexus System Interface }
\end{figure}
A secure authentication system is likely to be implemented for access to sensitive, organization-only functions such as getBalance. Although recipients would not have direct access to getBalance, they would receive transaction updates to ensure system security and transparency. Through the interface, users interact indirectly with the smart contract's functions, with actions executed on the backend. The interface may also include alerts for disbursements and data visualization tools for analyzing transaction trends. Therefore, the interface is a crucial link between users and the blockchain, facilitating accessibility for non-technical users.\newline

\subsection{System Functionality}

The "aNp" smart contract, written in Ethereum's Solidity programming language, provides the system's fundamental functionality. The contract stipulates the organization's address and mappings that maintain a list of recipients, their balances, and hashed bank account information for confidentiality. 
A constructor function assigns the contract-deploying address to the organization, thereby identifying the system administrator. Certain functions, restricted by the 'onlyOrganization' modifier, can be conducted only by this address, granting control over crucial tasks such as recipient management and fund distribution.
'addRecipient' and'removeRecipient' allow the organization to modify the recipient list, whereas'sendAllowance' enables the distribution of specified funds to authorized recipients, pending sufficient balance. 

The 'addFunds' function permits the organization to augment the Ether-based contract funds.
The 'getBalance' function allows any entity to check its balance, which is beneficial for recipients. The'registerBankAccount' function provides a secure method for storing the hashed bank information of the recipient, which is essential for off-chain fund transfers.
Key operations generate event logs ('AllowanceSent,' 'FundsAdded,' and 'BankAccountRegistered'), facilitating auditing and providing transparency. The contract encapsulates AidNeux's primary functionalities while emphasizing security, transparency, and effectiveness.

\section{Conclusion and Future Works} 
In conclusion, blockchain technology, and consortium blockchain in particular, has immense transformative potential across numerous industries. It provides a unique combination of transparency and privacy, addressing a number of extant issues. However, this emerging technology has limitations, notably scalability and interoperability problems. Future research should concentrate on addressing these obstacles while investigating innovative applications across industries. Moreover, as this technology permeates various industries, it may be necessary to develop new regulatory frameworks. This technological evolution and the regulatory environment will influence the future of consortium blockchain applications in an undeniable manner.


\begin{thebibliography}{00}

\bibitem{nb1} Al Sunaidi, S. J., \& Alhaidari, F. A. (2019). A survey of consensus algorithms for blockchain technology. In 2019 international conference on computer and information sciences (iccis) (pp. 1–6).

\bibitem{nb2} Onee, B. A. H., \& Antora, K. F., \& Rajme, O. S., \& Mansoor, N. (2023). Development of a Blockchain-Based On-Demand Lightweight Commodity Delivery System.  arXiv, cs.DC, 2307.08050.

\bibitem{nb3} Lashkari, B., \& Musilek, P. (2021). A comprehensive review of blockchain consensus mechanisms. IEEE Access, 9, 43620–43652.

\bibitem{nb4} Kim, S.-K., Kim, U.-M., \& Huh, J.-H. (2019). A study on improvement of blockchain application to overcome vulnerability of iot multiplatform security. Energies,
12(3), 402.

\bibitem{nb5} Yang, G., Lee, K., Lee, K., Yoo, Y., Lee, H., \& Yoo, C. (2022). Resource analysis of
blockchain consensus algorithms in hyperledger fabric. IEEE Access, 10, 74902–
74920.

\bibitem{nb6} Rahman, T., \& Mouno, S. I., \& Raatul, A. M., \& Azad, A. K. A., \& Mansoor, N. (2023). Verifi-Chain: A Credentials Verifier using Blockchain and IPFS. arXiv preprint arXiv: 2307.05797.

\bibitem{nb7} Bach, L. M., Mihaljevic, B., \& Zagar, M. (2018). Comparative analysis of blockchain
consensus algorithms. In 2018 41st international convention on information and
communication technology, electronics and microelectronics (mipro) (pp. 1545–1550).

\bibitem{nb8} Bai, Y., Hu, Q., Seo, S.-H., Kang, K., \& Lee, J. J. (2021). Public participation
consortium blockchain for smart city governance. IEEE Internet of Things Journal,
9(3), 2094–2108.

\bibitem{nb9} Hao, Y., Li, Y., Dong, X., Fang, L., \& Chen, P. (2018). Performance analysis of con-sensus algorithm in private blockchain. In 2018 ieee intelligent vehicles symposium
(iv) (pp. 280–285).

\bibitem{nb10} Baliga, A. (2016). The blockchain landscape. Persistent Systems, 3(5).
Fan, M., \& Zhang, X. (2019). Consortium blockchain based data aggregation and
regulation mechanism for smart grid. IEEE Access, 7, 35929–35940

\bibitem{nb11} Fu, Z., Dong, P., Li, S., \& Ju, Y. (2021). An intelligent cross-border transaction system based on consortium blockchain: A case study in shenzhen, china. Plos one, 16(6),E0252489.

\bibitem{nb12} Al-Shaibani, H., Lasla, N., \& Abdallah, M. (2020). “Consortium blockchain-based
decentralized stock exchange platform” IEEE Access, 8, 123711–123725

\bibitem{nb13} Kosba, A., Miller, A., Shi, E., Wen, Z., \& Papamanthou, C. (2016). Hawk: The
blockchain model of cryptography and privacy-preserving smart contracts. In
2016 ieee symposium on security and privacy (sp) (pp. 839–858).

\bibitem{nb14} Neisse, R., Steri, G., \& Nai-Fovino, I. (2017). A blockchain-based approach for data
accountability and provenance tracking. In Proceedings of the 12th international
conference on availability, reliability and security (pp. 1–10).

\bibitem{nb15} Zhang, X., \& Chen, X. (2019), Data security sharing and storage based on a
consortium blockchain in a vehicular ad-hoc network. IEEE Access, 7, 58241–
58254.

\bibitem{nb16} Purohit, S., Calyam, P., Alarcon, M. L., Bhamidipati, N. R., Mosa, A., \& Salah, K.
(2021). Honestchain: Consortium blockchain for protected data sharing in health
information systems. Peer-to-peer Networking and Applications, 14(5), 3012–3028.

\bibitem{nb17} Wang, D., \& Zhang, X. (2020). Secure data sharing and customized services for
intelligent transportation based on a consortium blockchain. IEEE Access, 8,
56045–56059.

\bibitem{nb18} Wei, G., \& Ma, Y. (2021). Privacy protection strategy of vehicle-to-grid network based on consortium blockchain and attribute-based signature. In Iop conference series: Earth and environmental science (Vol. 661, p. 012027).

\bibitem{nb19} Srivastav, R. K., Agrawal, D., Shrivastava, A., et al. (2020). A survey on vulnerabili-
ties and performance evaluation criteria in blockchain technology.

\bibitem{nb20} Akter, S., \& Mansoor, N. (2020). A spectrum aware mobility pattern based routing protocol for CR-VANETs. 2020 IEEE Wireless Communications and Networking Conference (WCNC), 1--6.

\bibitem{nb21} Mansoor, N., \& Hossain, M. I., \& Rozario, A., \& Zareei, M., \& Arreola, A. R. (2023). A Fresh Look at Routing Protocols in Unmanned Aerial Vehicular Networks: A Survey.IEEE Access.

\bibitem{nb22} She, W., Gu, Z.-H., Lyu, X.-K., Liu, Q., Tian, Z., \& Liu, W. (2019). Homomorphic
consortium blockchain for smart home system sensitive data privacy preserving.
IEEE Access, 7, 62058–62070.

\bibitem{nb23} Gu, J., Sun, B., Du, X., Wang, J., Zhuang, Y., \& Wang, Z. (2018). Consortium
blockchain-based malware detection in mobile devices. IEEE Access, 6, 12118–12128.

\bibitem{nb24} Xu, B., Xu, L. D., Wang, Y., \& Cai, H. (2022). A distributed dynamic authorisation
method for internet+ medical \& healthcare data access based on consortium
blockchain. Enterprise Information Systems, 16(12), 1922757.

\bibitem{nb25} Qiang, Z., Wang, Y., Song, K., \& Zhao, Z. (2021). Mine consortium blockchain: the
application research of coal mine safety production based on blockchain. Security
and Communication Networks, 2021.

\bibitem{nb26} Andrianova, A., \& Hauner, P., \& Mcdoland, D. K., \& Manning, D. A., \& Zerouali, M. (2022). Akropolis: A global blockchain pensions infrastructure.

\bibitem{nb27} Peng, S., Bao, W., Liu, H., Xiao, X., Shang, J., Han, L., Xu, Y. (2023). A peer-to-peer file storage and sharing system based on consortium blockchain. Future Generation Computer Systems, 141, 197–204.

\end{thebibliography}
\end{document}